\documentclass[conference]{IEEEtran}
\IEEEoverridecommandlockouts
\usepackage{caption}
\usepackage{subcaption}
\usepackage{graphicx}
\usepackage[table, svgnames, dvipsnames]{xcolor}
\usepackage{booktabs,makecell}
\usepackage{comment}
\usepackage{cite}
\usepackage{amsmath,amssymb,amsfonts}
\usepackage{algorithmicx}
\usepackage{algcompatible}
\definecolor{green}{rgb}{0.16, 0.67, 0.53}
\usepackage{textcomp,url}
\usepackage[left=1.62cm,right=1.62cm,top=1.9cm]{geometry}
\definecolor{chestnut}{rgb}{0.97, 0.51, 0.47}
\def\BibTeX{{\rm B\kern-.05em{\sc i\kern-.025em b}\kern-.08em
    T\kern-.1667em\lower.7ex\hbox{E}\kern-.125emX}}

\usepackage{algorithm}
\usepackage{algpseudocode}
\usepackage{dblfloatfix}
\usepackage{fixltx2e}

\usepackage{capt-of}

\definecolor{blue}{rgb}{0.0, 0.47, 0.75}

\usepackage{url}
\usepackage[hidelinks]{hyperref}

\begin{document}

\title{Energy-Aware Receiver-initiated MAC Protocol for Energy Harvesting IoT}

\author{
\IEEEauthorblockN{\IEEEauthorrefmark{2}Sohail Sarang, \IEEEauthorrefmark{3}Praveen Kumar Donta, and \IEEEauthorrefmark{2}Dejan Vukobratović}

\IEEEauthorblockA{\IEEEauthorrefmark{2}Faculty of Technical Sciences, University of Novi Sad, Trg Dositeja Obradovića 6, 21000 Novi Sad 21000, Serbia\\
\IEEEauthorrefmark{3}{Department of Computer Systems and Sciences}, {Stockholm University,} Stockholm 164 25, Sweden.} 
 \IEEEauthorrefmark{2}\{sohail, dejanv\}@uns.ac.rs, \IEEEauthorrefmark{3}praveen@dsv.su.se  }

\maketitle

\begin{abstract}
Energy harvesting (EH) technology has emerged as a key approach for extending the operational autonomy of Internet of Things (IoT) devices. In this context, medium access control (MAC) protocols for EH-IoT make use of the harvested energy efficiently to enhance the application performance. However, many existing MAC protocols rely on stored energy the battery in devising energy-aware strategies, which can interrupt the device operation and hinder the execution of ongoing tasks in dynamic and low EH conditions. To address this, this paper presents an energy-aware receiver-initiated MAC protocol for EH-IoT devices, called EARI-MAC, that adjusts the duty cycle of the receiver device based on stored energy and current harvesting power using the mathematical formulation. This approach supports sustained receiver operation under low EH conditions and improved application performance when harvested energy is abundant. The performance of the EARI-MAC has been evaluated in simulation using the real solar data gathered in Novi Sad, for three consecutive days. The simulation results demonstrate that the EARI-MAC utilizes the available energy resources efficiently and improves the packet delivery ratio and network throughput by more than 10\%, reduces the receiver energy consumption by 5.6\% and 8.2\%  and energy consumption per bit by 9.6\% and 13.9\% compared with QPPD-MAC and energy-unaware MAC, respectively. 
\end{abstract}


\section{Introduction}\label{sec:Introduction}
Internet of Things supports continuous monitoring of physical and environmental conditions in real-world applications, such as water quality monitoring, industrial process monitoring and others. IoT devices send the sensed data to the base station for real time analysis \cite{1huentenao2026multiparametric}. It is expected that there will be number 29.4 billion globally by 2030 \cite{2yu2022automating}. Most IoT devices currently rely on batteries, making their growing deployment costly to maintain and environmentally burdensome due to frequent battery replacement and disposal. Energy harvesting (EH) uses ambient solar, radio frequency, wind, and thermal sources to reduce battery dependence and environmental impacts, with surplus energy supporting improved application performance. However, the availability of energy is dependent on energy source and its surrounding conditions. For instance, solar energy is only available during the daytime. On rainy days, devices may not harvest enough energy and may stop working for long periods, which is undesirable for overall performance. The dynamic EH conditions also affect protocol level decisions, such as transmission scheduling, wake-up time, resource allocation and synchronization among devices. To address these challenges, energy-aware management schemes are crucial for reliable operation, as they adjust device activity, channel access probability, traffic load, relay selection, and sensing intervals according to the available energy \cite{3ashraf2025evaluating, 4oueis2016energy, 5yoo2011dynamic}.

Existing energy-aware medium access control (MAC) protocols of EH-IoT devices have used different approaches to utilize the harvested energy to improve the network performance. Ahmed et al. \cite{6ahmad2026adaptive} developed a framework that combines solar, thermal, and vibration energy sources to increase the available energy budget, supporting dynamic resource management and adaptive scheduling to improve network performance. QPPD-MAC presented in \cite{7sarang2018qos} supports different priority levels of data packets and uses the remaining  energy to adjust the receiver duty cycle. However, its reliance solely on stored energy limits its ability to sustain receiver operation during prolonged periods of low EH, potentially causing service interruptions. Gerndt et al. in \cite{8gerndt2025energy} presented an energy-aware duty cycle management technique that uses the weather prediction information to adapt device activity to guarantee continuous operation. Similarly, an adaptive duty cycling approach \cite{9ramkumar2025adaptive} was introduced that sets the duty cycle of the device based on residual stored energy, predicted harvested energy, and current traffic load. However, retrieving weather forecasts and estimating future harvested energy introduce communication and computation overhead, which may offset part of the energy savings in resource-constrained devices. An adaptive energy-management scheme presented in \cite{1huentenao2026multiparametric} that adjusts the duty cycle by switching among three fixed sleep durations, based on battery voltage and panel voltage. However, using the fixed sleep durations may limit its ability to make any adjustment to device activity in dynamic energy conditions.  The protocol presented in \cite{10hmidi2023new} follows the receiver initiated paradigm that combines both EH and device mobility aspects to ensure sustainable network operation. However, the performance of the protocol under varying EH conditions remains unexplored. An improved harvested energy management mechanism developed in \cite{11bengheni2026improved} adjusts device duty cycles based on residual energy, but its small-scale evaluation with predefined harvesting patterns leaves its robustness under unpredictable real-world EH conditions unverified. These limitations motivate the development of energy-aware protocols that consider both stored energy and harvesting variability to sustain long-term device operation and maintain network performance. 

To address this need, our work presents EARI-MAC, an energy-aware receiver-initiated MAC protocol that uses both stored battery energy and current harvesting power to adjust the duty cycle of receiver device through a mathematical formulation. The proposed protocol is evaluated using locally measured solar data with high resolution of 1s over three consecutive days \cite{12sarang2022solar}, and compared with existing EH-aware and energy-unaware protocols in terms of packet delivery ratio (PDR), network throughput, energy consumption per receiver, and energy consumption per bit.

\section{EARI-MAC Protocol}
\subsection{Communication overview }
\begin{figure}[b]
    \centering
    \includegraphics[width=0.49\textwidth]{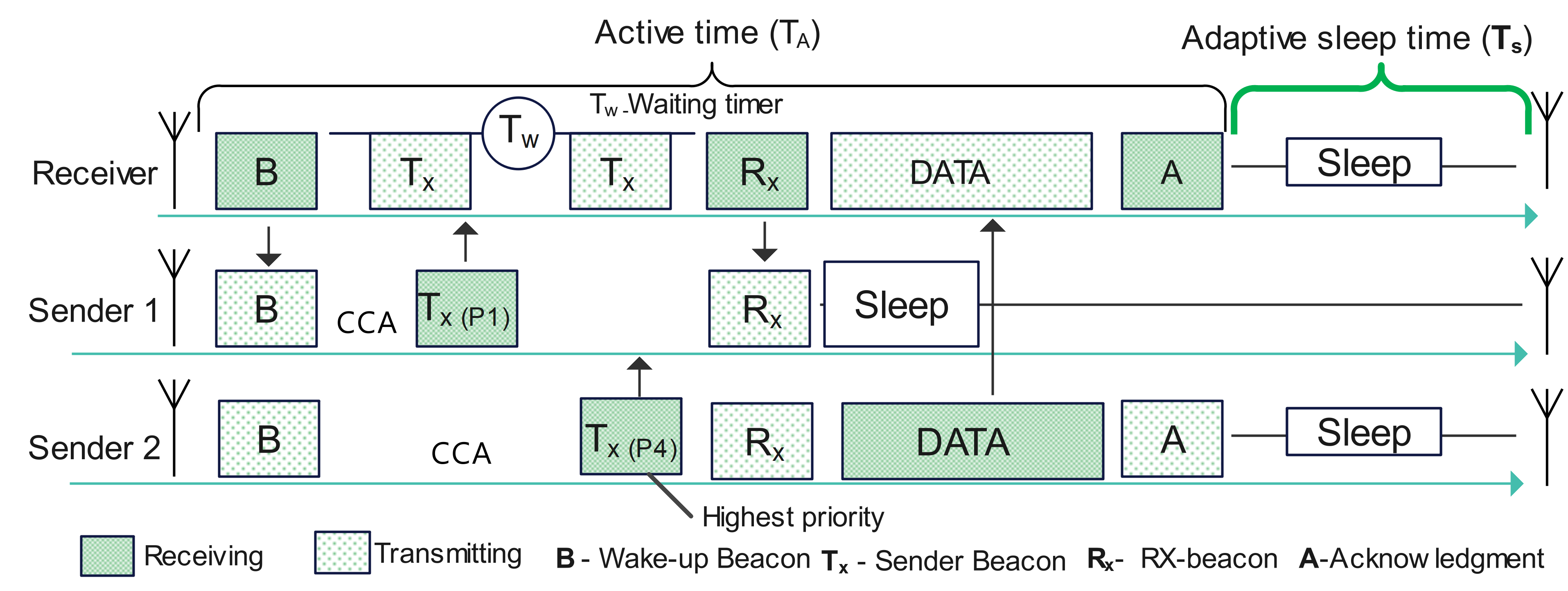}
    \caption{Communication overview in EARI-MAC}
    \label{fig1}
\end{figure}

This section presents an overview of communication between senders and receiver devices in EARI-MAC and describes the energy-aware duty cycle adjustment approach. EARI-MAC is a receiver-initiated protocol that follows a communication mechanism similar to QPPD-MAC, as shown in Fig. \ref{fig1}. At the beginning of each cycle, the receiver wakes up and initiates communication by broadcasting a wake-up beacon. It then waits for a duration ($T_w$) to receive transmission requests from the sender devices. The value of $T_w$ can be configured according to the number of senders. The sender devices with pending data packets listen to the channel for the wake-up beacon. Upon receiving it, they perform carrier sensing and transmit short transmission request beacons. Each request beacon contains the sender’s ID and packet priority. The priority level depends on the packet type and application requirements. After collecting the transmission requests beacons during $T_w$, the receiver selects a sender based on highest priority transmission request beacon received. It announces the selection by broadcasting an Rx beacon containing the selected sender’s ID. Upon receiving this beacon, the selected sender transmits its data packet, while the other senders enter sleep mode. This approach reduces idle listening and conserves energy at the non-selected senders, which can contend again in the next cycle. After successfully receiving the data packet, the receiver sends an acknowledgement to the selected sender.

\subsection{Energy-aware duty cycle adjustment }
Varying EH conditions drive the development of an adaptive duty cycle mechanism to sustain device operation and improve application performance in low and high EH scenarios, respectively. To utilize the external energy efficiently in both scenarios,  the receiver in EARI-MAC adjusts its duty cycle ($D_c$) based on the remaining energy level and current harvesting power, as follows
\begin{equation}\label{equation1}
     D_c  = D_{min}  + \left(D_{max}  - D_{min}\right)[\alpha e  + (1 - \alpha)h ]         
\end{equation}

where $D_{min}$   and $D_{max}$  denote the minimum and maximum duty cycles, respectively. The value of  $D_{min}$ can be set according to EH conditions. The variables e  and  $h$  represent normalized remaining energy and harvesting power, respectively. The weighting parameter, $\alpha \in [0, 1]$ controls the relative contributions of $e$ and $h$  in adjusting the duty cycle value. A larger $\alpha$ places greater emphasis on stored energy, whereas a smaller $\alpha$ gives greater weight to current harvesting power. The normalized remaining energy is calculated as
\begin{equation}\label{equation2}
    e=\frac{E_r}{E_{max}} 
\end{equation}
where $E_r$ and $E_{max}$  denote the remaining battery energy and maximum battery capacity, respectively. A threshold level (10\% of $E_{max}$) is applied to ensure that the battery does not completely deplete. 

The normalized harvesting power is given as
\begin{equation}\label{equation3}
    h =  \frac{H_p}{H_{(p,max)} }  
\end{equation}
where $H_p$ and $H_{(p,max)}$ denote the current harvesting power  and the maximum measured reference harvesting power, respectively. Then, the computed $D_c$ value and active time  $T_A$ are used to calculate the sleep time ($T_s$) of the receiver as shown in Eq.(\ref{equation4}). 
\begin{equation}\label{equation4}
    T_s=  \frac{(T_A\times (1-D_c))}{D_c} 
\end{equation}

\section{Performance Evaluation}

\subsection{Experimental and simulation setup}
\begin{figure}[t]
    \centering
    \includegraphics[width=0.45\textwidth]{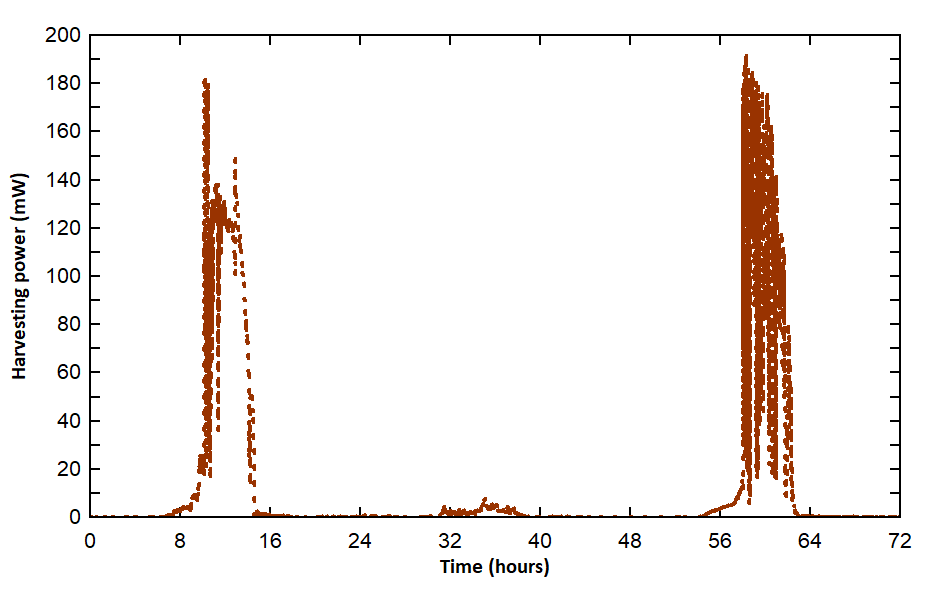}
    \caption{Harvesting power for three consecutive days \cite{12sarang2022solar}}
    \label{fig2}
\end{figure}

The performance of EARI-MAC is evaluated in a single-hop scenario in GreenCastalia \cite{13benedetti2013greencastalia}. Fig. \ref{fig2} presents the harvested power derived from solar irradiance measurements collected in Novi Sad, Serbia, from 5 to 8 January 2021 under varying EH conditions \cite{12sarang2022solar}. The solar data measurement setup included a Kipp \& Zonen CMP3 pyranometer connected to a DT80 data logger from Thermo Fisher Scientific. The simulations are performed for 72 consecutive hours, and its performance is compared with QPPD-MAC and energy-unaware receiver-initiated protocols. The receiver device in the energy-unaware protocol operates at a fixed duty cycle of 0.5. In the simulations, we used the parameters of the IXOLAR KXOB201K04F solar cell, which has an area of 9.66 cm$^2$ and a conversion efficiency of 25\%. The IoT device was modelled using the parameters of the TelosB platform and CC2420 radio. In this work, the maximum battery capacity, minimum duty cycle, maximum measured reference harvesting power, active time,  packet size,  and packet generation rate were set to 3024 Joules, 0.0000566, 182.78 mW, 17 ms, 28 bytes and 0.1 packets/s, respectively.
\begin{figure}[t]
    \centering
    \includegraphics[width=0.45\textwidth]{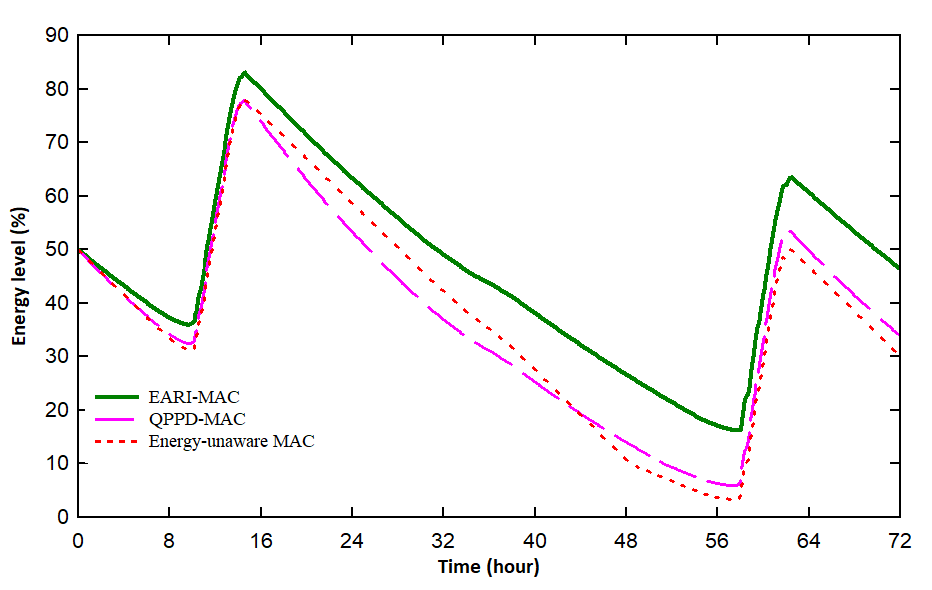}
    \caption{Receiver energy level when the number of sender devices is 7}
    \label{fig3}
\end{figure}
\begin{figure}[t]
    \centering
    \includegraphics[width=0.45\textwidth]{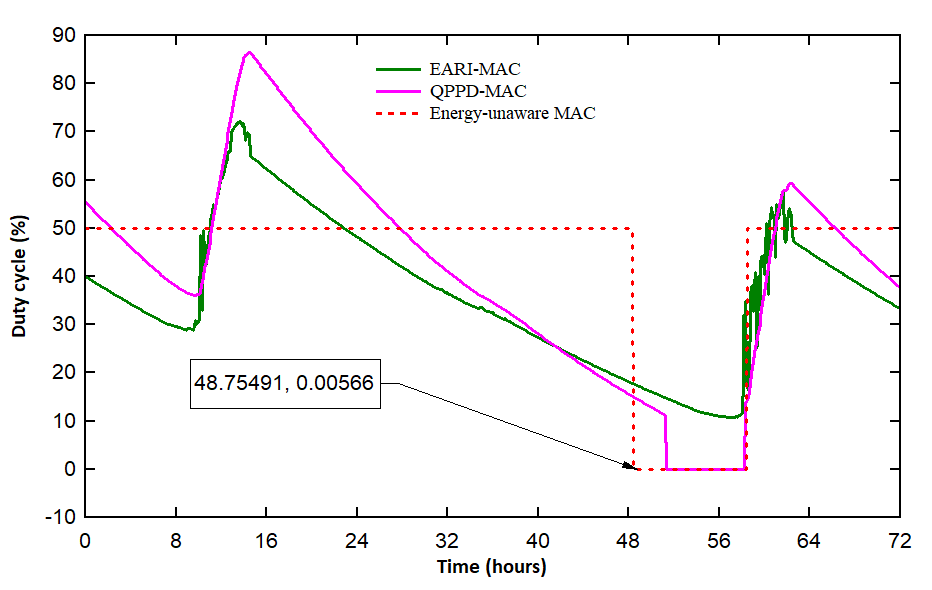}
    \caption{Receiver duty cycle when the number of sender devices is 7}
    \label{fig4}
\end{figure}
\begin{figure}[t]
    \centering
    \includegraphics[width=0.45\textwidth]{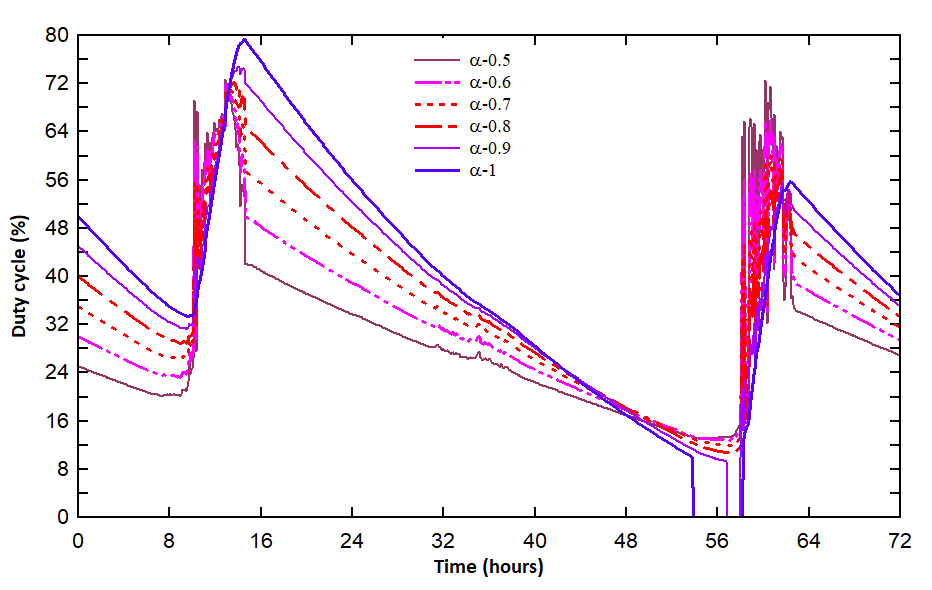}
    \caption{The influence of the weighting parameter, $\alpha$ on duty cycle of the receiver}
    \label{fig5}
\end{figure}
\subsection{Results and discussion}
Fig. \ref{fig3} presents the remaining energy of the receiver as a percentage of battery capacity for EARI-MAC, QPPD-MAC, and Energy-unaware MAC. In all three protocols, the receiver starts with an energy level of 50\%. At the end of the first day, the remaining energy levels are 58.4\%, 51.3\%, and 56.8\%, respectively. The proposed EARI-MAC has more stored energy compared to other two protocols. During the second day, due to limited solar energy availability, the energy levels decrease to 21.5\% for  EARI-MAC, 12.9\% for QPPD-MAC, and 9.5\% for Energy-unaware MAC. Energy-unaware MAC consequently falls below the operational threshold of 10\%. Approximately three hours later, QPPD-MAC also reaches this threshold. As a result, the receivers in both QPPD-MAC and Energy-unaware MAC remain nonoperational for almost 7 and 10 hours, respectively, whereas EARI-MAC maintains sufficient energy to continue operating. On the third day, the energy levels increase and reaches to 59.1\%, 53.4\%, and 49\% for EARI-MAC, QPPD-MAC, and Energy-unaware MAC, respectively, before declining to 41.7\%, 34.2\%, and 30.2\% by the end of the observation period. The duty cycle of the receiver in all protocols are shown in Fig. \ref{fig4}. The proposed EARI-MAC adjusts its duty cycle using both the remaining energy and the current harvesting power using the Eq. (\ref{equation1}), that allows the receiver to respond to changes in energy availability effectively. The weighting parameter $\alpha$ controls the relative contribution of the remaining energy level and harvesting power in adjusting the duty cycle. Different values of $\alpha$ were investigated to assess the influence of harvesting power on the receiver  duty cycle of the receiver in EARI-MAC, as shown in Fig. \ref{fig5}. Based on this investigation, $\alpha = 0.8$ was selected for this work to sustain receiver operation and achieve improved network performance.
\begin{figure}[t]
    \centering
    \includegraphics[width=0.45\textwidth]{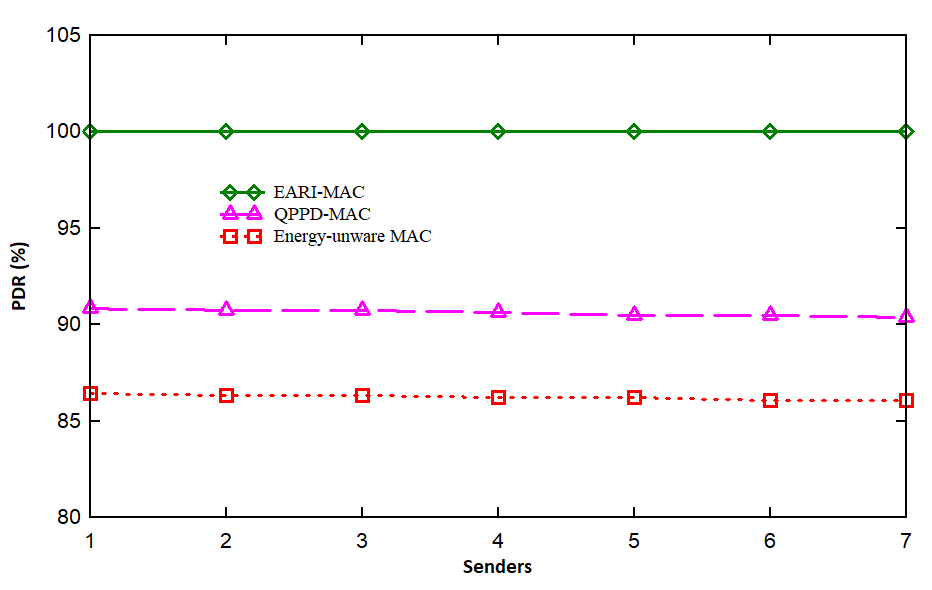}
    \caption{PDR  with varying number of sender devices}
    \label{fig6}
\end{figure}
\begin{figure}[t]
    \centering
    \includegraphics[width=0.45\textwidth]{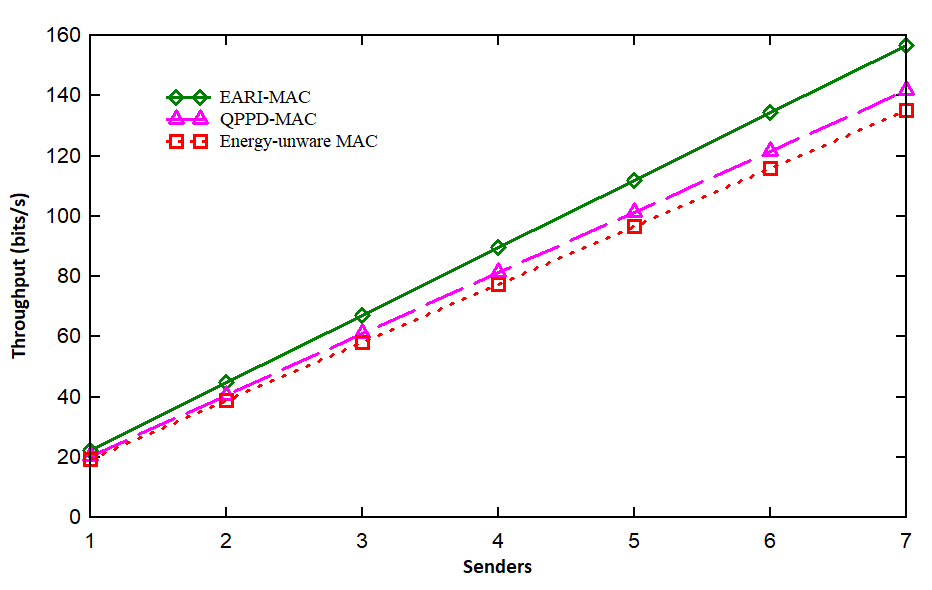}
    \caption{Network throughput  with varying number of sender devices}
    \label{fig7}
\end{figure}
\begin{figure}[t]
    \centering
    \includegraphics[width=0.45\textwidth]{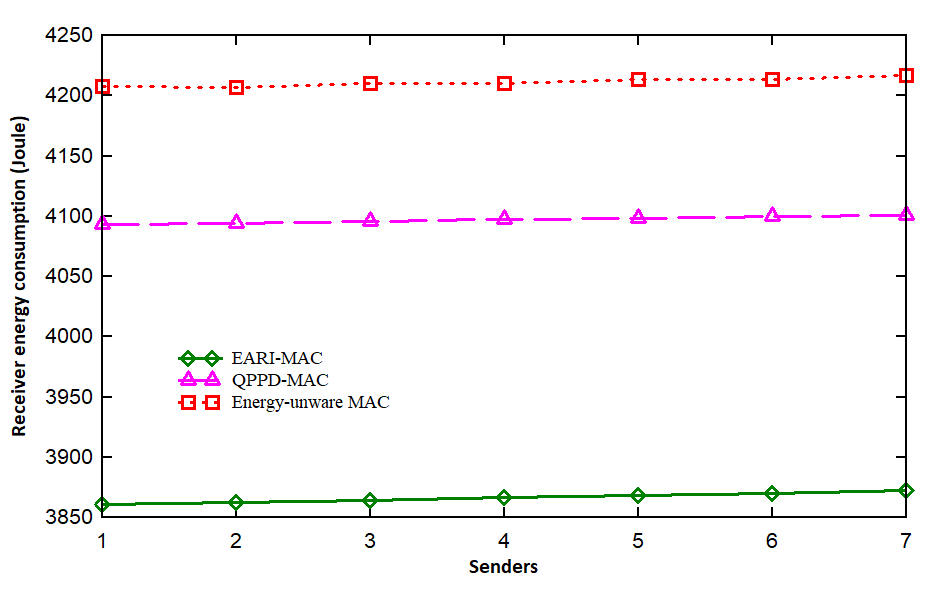}
    \caption{Energy consumption of receiver versus number of sender devices}
    \label{fig8}
\end{figure}
\begin{figure}[t]
    \centering
    \includegraphics[width=0.45\textwidth]{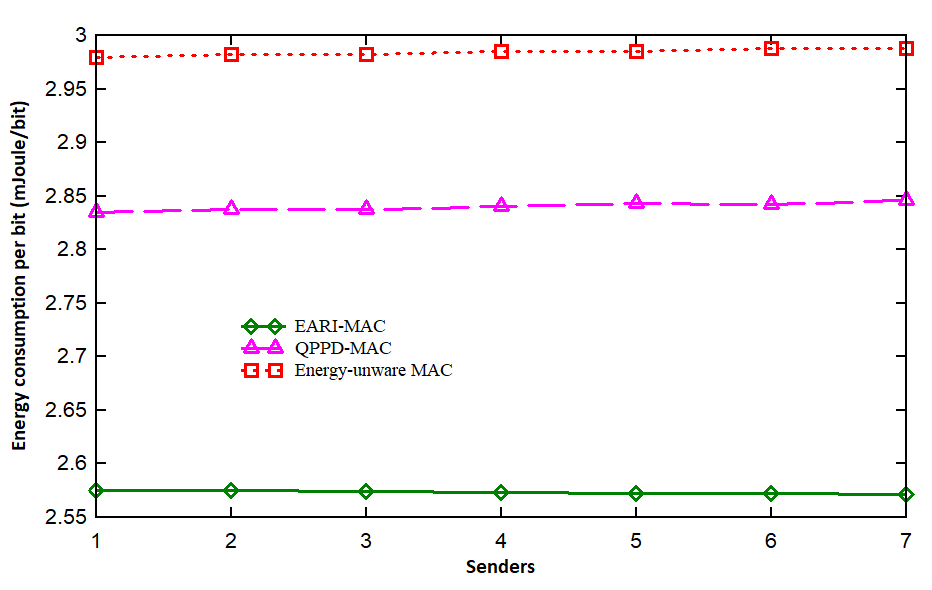}
    \caption{Energy consumption per bit versus number of sender devices}
    \label{fig9}
\end{figure}

Fig. \ref{fig6} compares the PDR of EARI-MAC with that of QPPD-MAC and the energy-unaware MAC protocol. EARI-MAC maintains 100\% PDR as the number of senders increases from 1 to 7. In contrast, the PDR of QPPD-MAC and Energy-unaware MAC decline to 90.4\% and 86.1\%, respectively, with 7 senders. EARI-MAC therefore achieves improvements of 9.6\% and 13.9\%  when compared to other two protocols. The reason behind is that in EARI-MAC the receiver considers both the remaining energy level and the current harvesting power to set its duty cycle which enables continuous operation throughout the three consecutive days. In other protocols, due to the disruption in operation of the receiver in QPPD-MAC and Energy-unaware MAC, the  PDR decreases significantly  due to operational interruption. EARI-MAC also achieves more than 10\% higher throughput than the other protocols across all evaluated numbers of sender devices, as shown in Fig. \ref{fig7}. 

Fig. \ref{fig8} presents the receiver energy consumption for all three protocols. The energy consumption increases slightly as the number of senders rises from 1 to 7, because the receiver receives and processes more data packets. The proposed EARI-MAC consumes less receiver energy than QPPD-MAC and Energy-unaware MAC and it reduces receiver energy consumption by 5.6\% and 8.2\% compared to other protocols, respectively. Furthermore, EARI-MAC also achieves the lowest energy consumption per delivered bit across all sender devices, as shown in Fig. \ref{fig9}. EARI-MAC reduces energy consumption per bit by 9.6\% and 13.9\% compared with QPPD-MAC and Energy-unaware MAC, respectively. In EARI-MAC, the energy-aware duty cycle adjustment approach helps the receiver to sustain its operation. As a result, it receives all packets transmitted by the senders. In contrast, the receivers in QPPD-MAC and Energy-unaware MAC protocols remain nonoperational for almost 7 and 10 hours, respectively. These operational disruptions in the operation reduces the number of successfully received packets and contribute to higher energy consumption per bit.

\section{Conclusion and Future work}
This paper presented EARI-MAC, an energy-aware receiver-initiated MAC protocol for EH-IoT devices that adjusts the receiver’s duty cycle based on stored battery energy and current harvesting power. The performance of EARI-MAC was evaluated through simulation using locally collected solar data over 72 hours and compared with the other two protocols. The performance evaluation considered PDR, network throughput, receiver energy consumption, and energy consumption per delivered bit. The simulation results show that EARI-MAC improved PDR and network throughput by more than 10\% compared to other protocols. Moreover, EARI-MAC reduced receiver energy consumption by 5.6\% and 8.2\% and energy consumption per bit by 9.6\% and 13.9\%, compared to QPPD-MAC and energy-unaware MAC, respectively. The work also investigated the influence of the weighting parameter on duty cycle of the receiver that can help to balance the energy consumption of the receiver under varying EH condition to achieve the desired application performance. Future work will focus on the design and implementation of an energy-aware MAC protocol for IoT devices that incorporates multiple EH sources.

\section*{Acknowledgment}
This work has received funding from the European Union’s Horizon Europe research and innovation programme MSCA Staff Exchanges 2025 under the grant agreement No 101299722.

\bibliographystyle{IEEEtran}
\footnotesize
\bibliography{ref.bib}
\end{document}